\documentclass[runningheads]{llncs}
\usepackage[T1]{fontenc}
\usepackage{graphicx}
\usepackage{amsmath} % Need to load amsmath before the documentclass is defined.
\usepackage{braket}
\usepackage{float}
\usepackage{subfig}
\usepackage{tikz}
\usetikzlibrary{quantikz2}

\usepackage{comment}
\begin{document}
\title{Emulation of Entanglement Distribution Networks on a Quantum Computer}
\author{Ashley N. Tittelbaugh\inst{1}\orcidID{0009-0003-9296-5706} \and
Jerry Horgan\inst{2}\orcidID{0000-0001-5783-6077} \and
Rohan Bali\inst{1}\orcidID{0009-0007-7992-8803} \and
Marco Ruffini\inst{2}\orcidID{0000-0001-6220-0065} \and
Daniel C. Kilper\inst{2}\orcidID{0000-0003-3542-5335} \and
Shelbi L. Jenkins\inst{1}\orcidID{0009-0003-6108-9790} \and
Boulat A. Bash\inst{1}\orcidID{0000-0002-1205-3906}
}
\authorrunning{A. Tittelbaugh et al.}
\institute{The University of Arizona, Department of Electrical and Computer Engineering, Tucson AZ 85721, USA 
\email{\{antittelbaugh,rbali,shelbijenkins,boulat\}@arizona.edu} \and
Trinity College Dublin, Dublin, Ireland
\email{\{horganj3,marco.ruffini,dan.kilper\}@tcd.ie}
}
\maketitle              % typeset the header of the contribution
\begin{abstract}
We investigate how quantum computers can be used to emulate quantum networks and study their performance under practical impairments.
In particular, we evaluate how degraded entanglement and communication latency affect teleportation-based distributed multipartite-entanglement-state construction.
We model imperfect Bell-pair sources using depolarizing noise channels and classical communication delays using thermal relaxation.
We implement the depolarization using Stinespring dilation, randomly applied Pauli errors, and quasi-probability decompositions, evaluating the latter two on IQM quantum hardware and all three in simulation.
We then study the performance of the entanglement distribution under noise generated by the aforementioned models.  
Although these noise models are mathematically equivalent, we find that hardware constraints result in profound differences in the corresponding results, highlighting the importance of careful experiment design. 

\end{abstract}

\section{Introduction}
Effective distributed quantum computing (DQC) depends on developing algorithms tailored to the physical limitations and network architectures of connected quantum processing units (QPUs). Current algorithmic design efforts generally pursue two main strategies. The first is circuit cutting, where a global quantum circuit is partitioned into smaller sub-circuits that can be executed in parallel on separate QPUs, and later classically recombined \cite{hofmann2009simulate,mitarai2021constructing,piveteau2023circuit,peng2020simulating,11131294}. The second is graph state-based design, which constructs large entangled graph states spanning multiple QPUs to support algorithms exceeding the qubit capacity of any single processor \cite{fan2025optimized}. While circuit cutting is easier in the short term, as it only requires classical post processing and can be implemented on one smaller QPU, it has a sampling overhead that is asymptotically ${O}\left(\gamma^{2n}\right)$, where $n$ is the number of circuit cuts~\cite{CarreraVazquez2024}. In contrast, if multiple QPUs or quantum computers are interconnected through network-distributed entanglement, the graph state approach could execute the entire algorithm simultaneously, eliminating this overhead \cite{CarreraVazquez2024}. However, because practical network-based entanglement distribution has not yet been achieved, it remains challenging to design and evaluate algorithms that depend on such large, distributed graph states.

Carrera Vazquez \textit{et al.}~\cite{CarreraVazquez2024}, propose a potential solution to this limitation through the use of cut Bell pairs, discussed in detail in Section~\ref{sec:cut Bell pair}. In their approach, the density matrix of a Bell pair is decomposed into a set of separable states, each of which can be classically combined in post-processing to reproduce the same measurement statistics as those obtained from an entangled Bell pair. Using a classical-communication-enabled teleportation protocol~\cite{piveteau2023circuit,CarreraVazquez2024}, they demonstrate that these cut Bell pairs can be teleported to create a virtual  controlled-Z (CZ) gate across a previously disconnected edge in a graph state, virtually entangling qubits that were initially unentangled~\cite{CarreraVazquez2024}. 

In order to be used as a tool for algorithmic development with the intention of having deployable algorithms for network-enabled distributed quantum computing systems, one must account for the constraints imposed by generation and distribution, namely the fidelity and rate degradation. Here, we focus on the fidelity, leaving investigation of rate to future work. Since we assume a zero-added-loss multiplexed (ZALM) entanglement source~\cite{Chen_ZALM} with a predicted emission fidelity of 99.6\%, we apply three distinct depolarizing channels to the cut Bell pairs which can be tuned to model any realistic input fidelity. These methods are described in depth in~\cite{tittelbaugh2025modeling}. Our performance analysis in simulation and on a quantum computer informs future modeling and design of network-enabled distributed quantum computing. 

\section{Cut Bell Pair}
\label{sec:cut Bell pair}
Cut Bell pairs consist of a set of carefully chosen separable states whose outputs, when linearly combined with appropriate weights, reproduce the same measurement statistics as an entangled $\ket{\Phi^+}$ Bell pair~\cite{CarreraVazquez2024}, using the density matrix decomposition of Vidal \textit{et al.}~\cite{vidal1999robustness}:
\begin{equation}
      \rho_k = (1 + t_k)\rho_k^{+} - t_k\rho_k^{-},
      \label{eq:ogQPD}
\end{equation}
where $k$ is the number of cut Bell pairs, $t_k = 2^k-1$, and $\rho_k^{\pm}$ are separable states. The signal term $\rho_k^{+}$ contains states with arbitrary single-qubit superpositions that remain separable across the bipartition~\cite{CarreraVazquez2024,vidal1999robustness}, while the corrective term $\rho_k^{-}$ is composed of basis states orthogonal to the target (e.g., $|01\rangle$ and $|10\rangle$ for $\ket{\Phi^+}$). The weights $a_{i,k}$ form a quasi-probability distribution (QPD), meaning they sum to~1 but may take negative values, with sampling overhead $\gamma_k = 2t_k + 1$. For a single cut Bell pair ($k{=}1$), $I_{\text{LOCC}} = n_k^- + n_k^+ = 5$ independent circuit realizations are required. We use the optimized circuit and parameters of Carrera Vazquez \textit{et al.}~\cite{CarreraVazquez2024}, summarized in Table~\ref{tab:rho_thetas}. Further details on the QPD construction, depolarization methods applied to the cut Bell pairs, and the entanglement witness formalism used in our evaluation are given in~\cite{tittelbaugh2025modeling}.

\begin{table}[htpb]
\centering

\renewcommand{\arraystretch}{1.2}
\begin{tabular}{c c c c c c c}
\hline
 & $  \theta_4[0]  $ & $\theta_4[1]  $ & $\theta_4[2]  $ & $\theta_4[3]  $& Basis State & $P_i  $\\ \hline
$\rho_k^{+}$ & $\tfrac{\pi}{2}$ & $0$ & $\tfrac{\pi}{2}$ & $0$ & - & $\tfrac{2}{9}$ \\
$\rho_k^{+}$ & $\tfrac{\pi}{2}$ & $-\tfrac{2\pi}{3}$ & $\tfrac{\pi}{2}$ & $\tfrac{2\pi}{3}$ &-  & $\tfrac{2}{9}$ \\
$\rho_k^{+}$ & $\tfrac{\pi}{2}$ & $\tfrac{2\pi}{3}$ & $\tfrac{\pi}{2}$ & $-\tfrac{2\pi}{3}$ & - &$\tfrac{2}{9}$  \\
$\rho_k^{-}$ & $\pi$ & $0$ & $0$ & $0$ & $|10\rangle$ & $\tfrac{1}{6}$ \\
$\rho_k^{-}$ & $0$ & $0$ & $\pi$ & $0$ & $|01\rangle$ & $\tfrac{1}{6}$ \\ \hline
\\[1ex]
\end{tabular}
\caption[Single-cut Bell pair parameters] {\textbf{Single-cut Bell pair parameters} Parameter sets, $\theta_4$, for one cut Bell pair. Each line of the table represents one distinct circuit run needed to reconstruct the Bell pair. Each $\rho_k^-$ represents a basis state that is orthogonal to the target Bell pair. The weights of the constructed probability distribution are given as $P_i$.}
\label{tab:rho_thetas}
\end{table}

\section{Experiment Setup}
\label{sec:setup}
\begin{figure}[htbp]
\centering
% Column 1: The Text Block (Takes up 55% of the text width)
\begin{minipage}{0.60\textwidth}
To evaluate how network-distributed entanglement performs when used to create an entanglement graph state, we create and evaluate the edge shown in Fig.~\ref{fig:cut_graph_state} as (0,5).

We transform a linear graph state~\cite{tittelbaugh2025modeling} into a ring graph state using the teleportation protocol~\cite{piveteau2023circuit,CarreraVazquez2024}. The generic overall circuit for this experiment is shown in Fig.~\ref{fig:cut graph state}. Notably, this generic circuit omits the depolarizing channels for readability. The depolarizing circuits, if applicable, are added before the teleportation protocol.

To calculate the entanglement witnesses~\cite{jungnitsch2011entanglement}, which provide a lower bound on the bipartite-entanglement fidelity, and consequently the fidelity, of each entangled qubit pair in the resulting ring graph state, the circuit is executed once for each required observable, $S_0$–$S_5$ and $S_{0,1}$–$S_{5,0}$, resulting in a total of 12 circuit runs. 
\end{minipage}
\hfill % Pushes the two blocks to the absolute edges
% Column 2: The Image Block (Takes up 40% of the text width)
\begin{minipage}{0.35\textwidth}
\centering
\includegraphics[width=\textwidth]{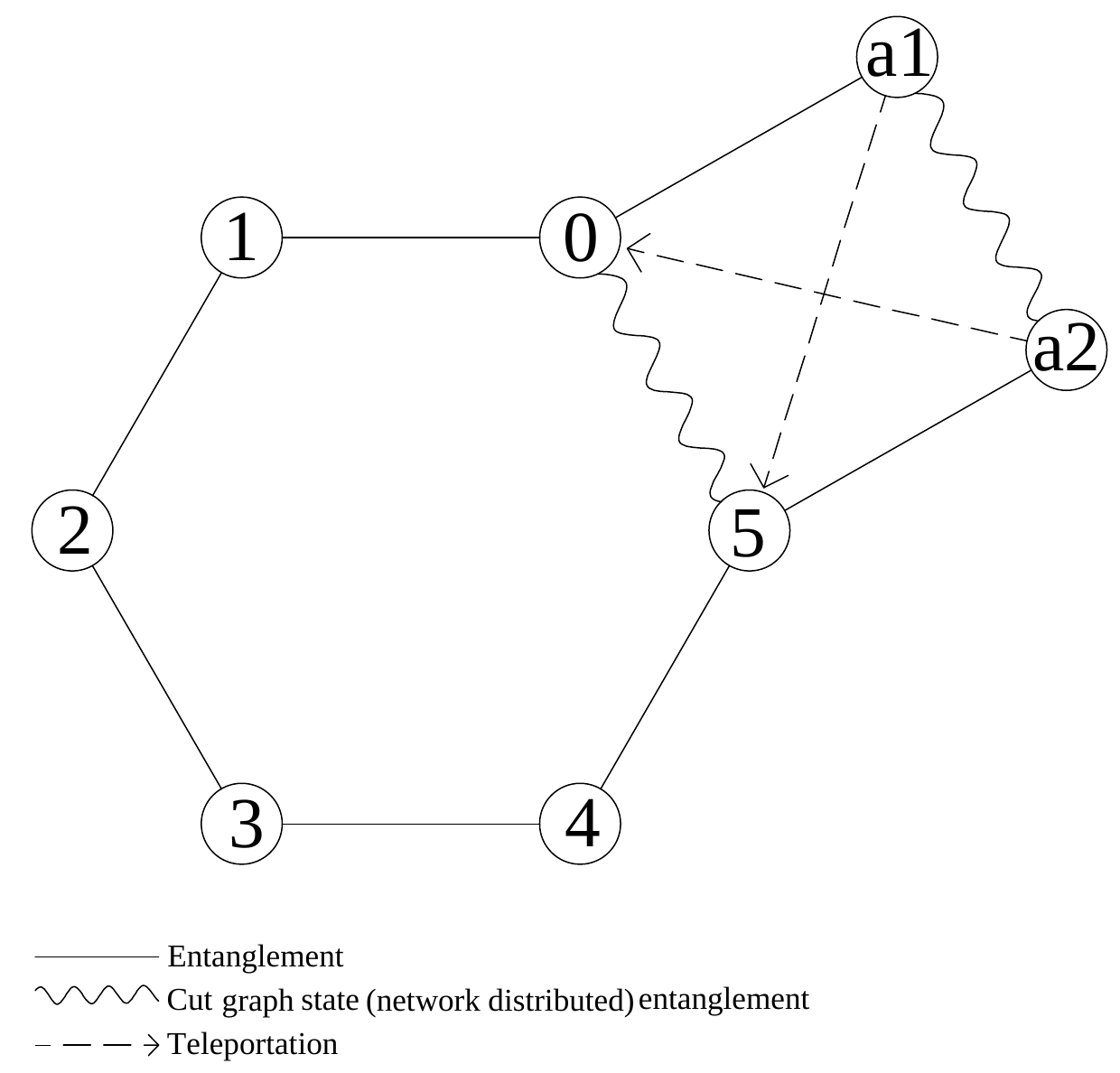} 
\caption{Ring graph state; edge~(0,5) is realized via a teleported virtual CZ using a cut Bell pair (Fig.~\ref{fig:cut graph state}).}
\label{fig:cut_graph_state}
\end{minipage}
\end{figure}

\begin{figure}[htbp]
\centering
\includegraphics[width=\columnwidth]{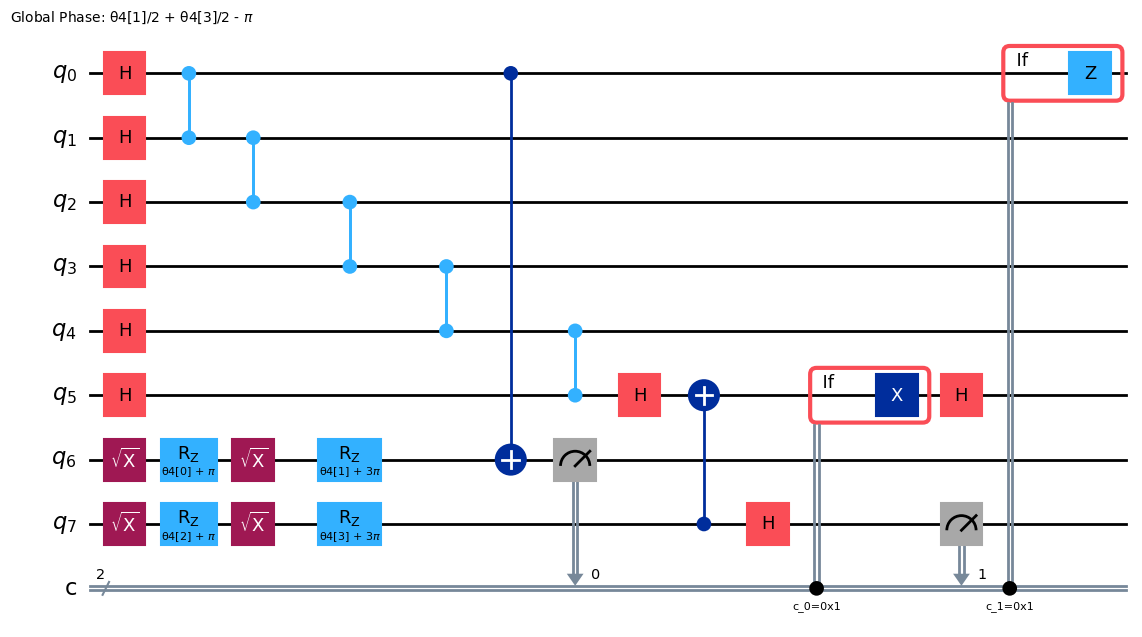}
\caption[Cut graph-state circuit] {\textbf{Cut graph-state circuit} The base form of the cut graph state circuit runs in all experiments. }
\label{fig:cut graph state}
\end{figure}

Each of these 12 circuits is executed for every term in the associated QPD decomposition, and subsequently combined using the QPD observable formula given in~\cite{CarreraVazquez2024}. Thus, for a single set of input parameters, a total of 60 circuits are executed (or 84 when using QPD-based depolarization~\cite{tittelbaugh2025modeling}). This number could potentially be reduced by measuring commuting stabilizers simultaneously, at the cost of more complex post-processing. 

We evaluate the efficacy of each of the three depolarizing methods discussed in~\cite{tittelbaugh2025modeling} in comparison to a cut Bell pair at ideal fidelity, i.e., not affected by a depolarizing channel. Table~\ref{tab:depol_methods} summarizes the relevant gate, qubit and circuit run differences between each method. Whenever possible, we evaluate using both a simulation based on IBM's Torino (a 133-qubit Heron r1 device~\cite{abughanem2024ibm}) and IQM's Emerald (a 54-qubit superconducting processor~\cite{aws_iqm_2025}).

\begin{table}[H]
\centering
\renewcommand{\arraystretch}{1.2}
\begin{tabular}{c c c c c}
\hline
  Method & Gates& Qubits & Runs per Observable& Described in \\ \hline
Ideal & - & - & - & Fig.~\ref{fig:cut graph state}  \\
Unitary & $\sim$ 60 & 4 & 0 & \cite{tittelbaugh2025modeling}  \\
Pauli & $\sim$ 10 & 6 & 0 & \cite{tittelbaugh2025modeling} \\
QPD & 0 & 0 & 2 & \cite{tittelbaugh2025modeling} \\
\hline
\\[1ex]
\end{tabular}
\caption[Depolarization methods] {\textbf{Depolarization methods} Depolarization methods as compared to ideal fidelity entangled pairs (no depolarization). Each of the columns is quantified as the additional gates, qubits, or runs per observable needed compared to ideal. }
\label{tab:depol_methods}
\end{table}
We evaluate the different depolarization methods assuming a ZALM emission fidelity of 99.6\% under a 10~ns classical communication delay. Because the simulation environment does not inherently account for classical communication time, we manually impose a fixed 10~ns delay, corresponding to a propagation distance of approximately 2 meters in optical fiber. In addition, we analyze the impact of network-distributed fidelity on the cut graph state by sweeping from 50\% to 100\%, in increments of 10\%, in both simulation and hardware experiments. Finally, we investigate the influence of classical communication delay by varying the simulation delay logarithmically from 0 to 100,000~ns. For each test configuration, we report the individual-qubit stabilizer expectations and the pairwise fidelities corresponding to the six-qubit ring graph state. Each circuit is executed with 1,024 shots.

\subsection{Simulation}
\label{sec:simulation}
In our experiments, we simulate the circuits using a noise model based on IBM Torino, a 133-qubit device built around the Heron r1 processor architecture \cite{abughanem2024ibm}. The noise model incorporates a snapshot of the gate errors reported for Torino at the start of the program, and this fixed model is held constant for all runs corresponding to the same set of input parameters. However, separate runs for different depolarization methods each retrieve an independent calibration snapshot, so the noise model may differ slightly between methods. Because the standard noise model does not account for classical communication latency, we explicitly insert a variable delay on all graph state qubits following measurement, and prior to the corresponding conditional feed-forward gates.

It is important to note that classical communication delays are the only point where we apply thermal relaxation, as we assume the time is dominated by classical communication.

\subsection{Quantum Computing Hardware}
\label{sec:hardware}
For experiments, we employ IQM's Emerald, a 54-qubit superconducting processor with a square lattice connectivity \cite{aws_iqm_2025}. The square lattice connectivity of the quantum computer does not provide needed connectivity to run the unitary depolarization on hardware in its current state. 

\section{Results and Discussion}
\label{sec:results}
\subsection{Simulated Results}
\label{sec:simulated results}
\begin{figure}[htpb]
\centering
\subfloat[ Simulated graph state stabilizer \\values]{\includegraphics[width=.48\columnwidth]{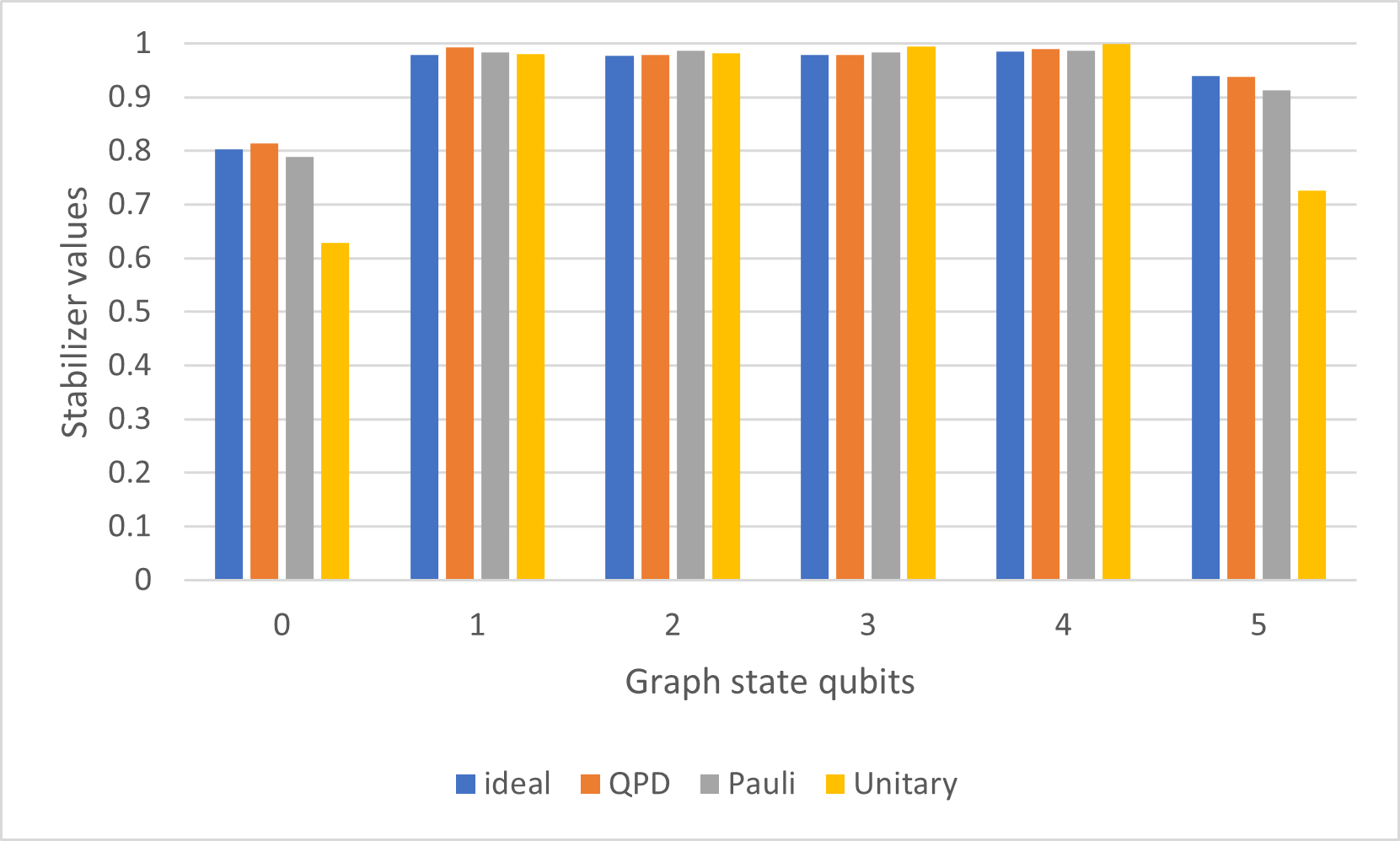} \label{fig:sim_standard_stabilizers}}
\subfloat[Simulated lower bound of bipartite-entanglement fidelity across graph state edges]{\includegraphics[width=.48\columnwidth]{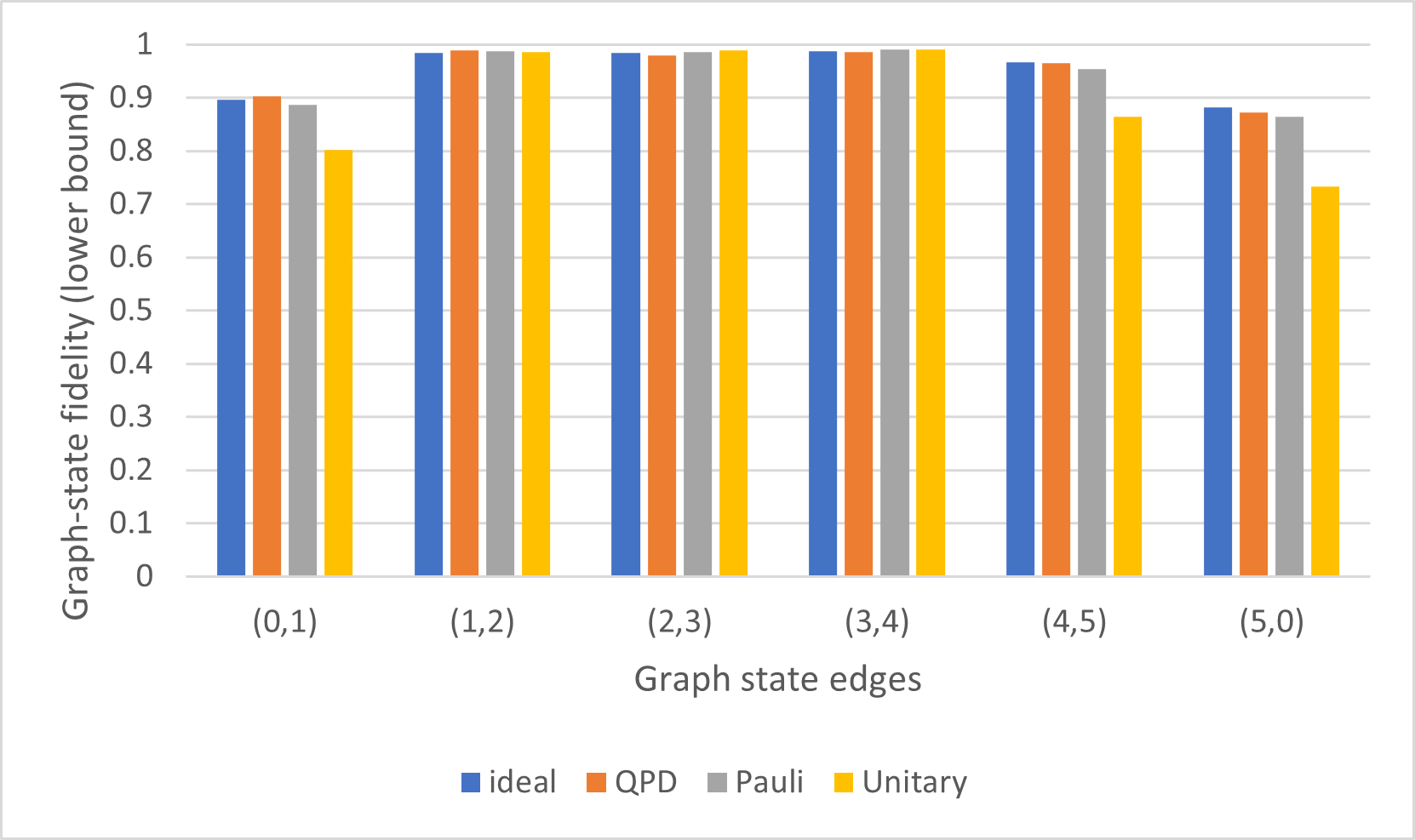} \label{fig:sim_standard_fidelity}}\\
\caption[Simulation at 99.6\% ZALM fidelity and 10~ns classical communication] {\textbf{Simulation at 99.6\% ZALM fidelity and 10~ns classical communication} Individual graph state stabilizer values and a lower-bound for pair-wise graph state fidelities. For both stabilizers and fidelity, a unity value indicates perfect entanglement with no errors. }
\label{fig:sim_standard}
\end{figure}

Fig.~\ref{fig:sim_standard} presents the simulation results assuming a 99.6\% ZALM-generated fidelity and a classical communication delay of 10~ns. We evaluate each of the depolarizing channels described in~\cite{tittelbaugh2025modeling}, along with the ideal case where no depolarization is applied. Consistent with the findings of Carrera Vazquez \textit{et~al.}~\cite{CarreraVazquez2024}, the lowest stabilizer values, shown in Fig.~\ref{fig:sim_standard_stabilizers}, occur at the qubits connected by the cut edge, namely qubits~0 and~5. We hypothesize that the lower fidelity observed on qubit~0 relative to qubit~5 arises from the virtual \texttt{cz} gate implemented during the teleportation process~\cite{piveteau2023circuit,CarreraVazquez2024}. In this configuration, qubit~5 functions as the control and qubit~0 as the target. Thus, any errors affecting qubit~5 propagate to qubit~0 but not vice versa.

Fig.~\ref{fig:sim_standard_fidelity} further illustrates that the lower-bounded fidelity across the cut edge (0,5) is the smallest, while the lower-bounded fidelities of neighboring edges (0,1) and (4,5) exhibit more moderate degradation. This behavior is again consistent with the results of Carrera Vazquez \textit{et~al.}~\cite{CarreraVazquez2024}. Mirroring the stabilizer behavior, edge~(0,1) exhibits a lower fidelity than its counterpart edge~(4,5), which we similarly attribute to the directionality of the \texttt{cz} operation.

Depolarization differences are most pronounced near the cut edge, where unitary depolarization performs notably worse due to its additional gate overhead. QPD and Pauli perform close to ideal, with Pauli slightly lower due to increased gate errors. All methods maintain fidelities above~0.5, confirming entanglement across the cut. Occasional instances where a depolarized run outperforms the ideal case arise because each method is executed as a separate run, each retrieving a fresh calibration snapshot from the hardware backend, so the effective noise model differs slightly between methods.
  
\subsection{Hardware Run Results}
\label{sec:hardware results}
\begin{figure}[htpb]
\centering
\subfloat[Graph state stabilizer values on\\ hardware]{\includegraphics[width=.5\columnwidth]{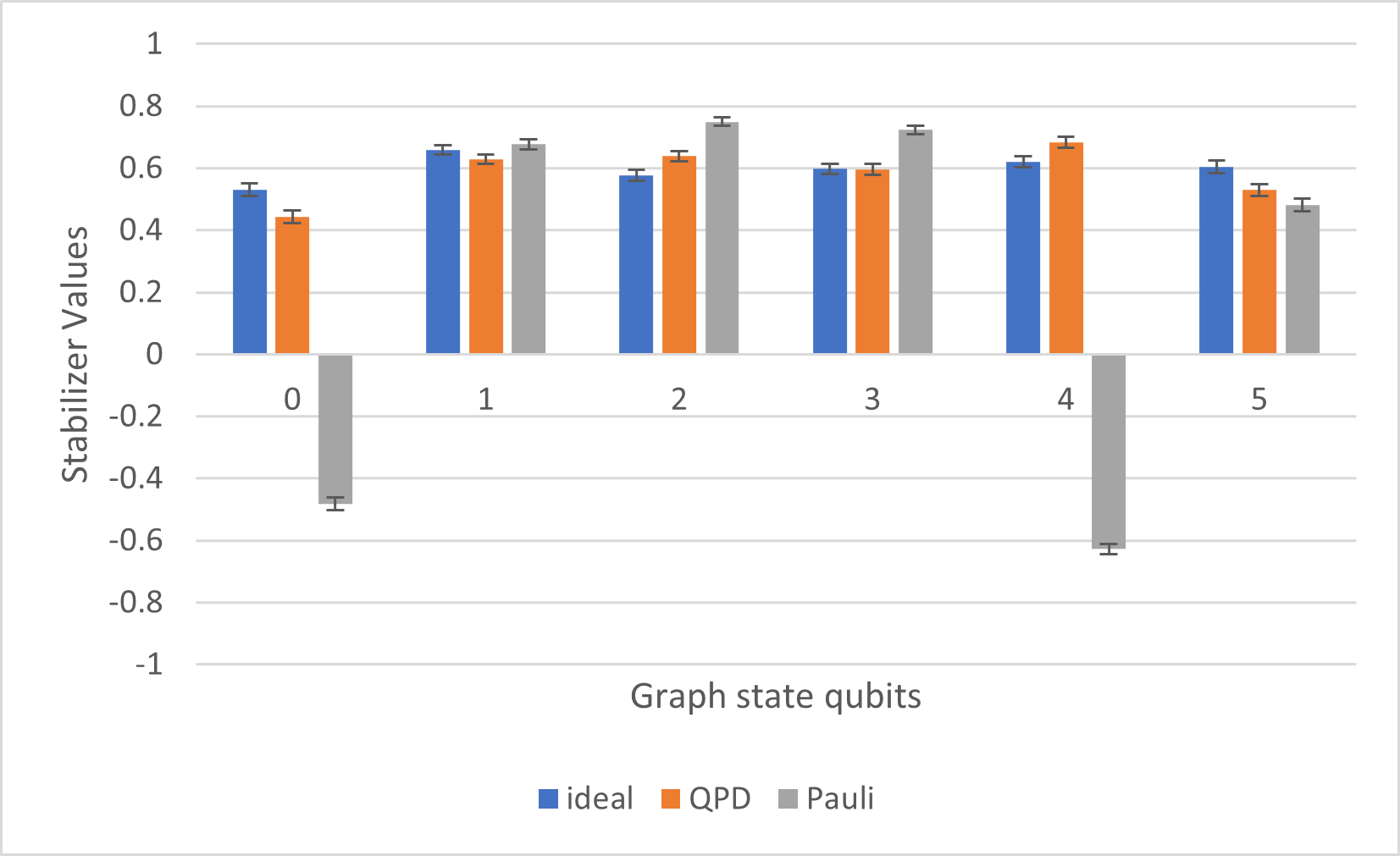} \label{fig:real_standard_stabilizers}}
\subfloat[Lower bound for bipartite-entanglement fidelity across graph state edges on hardware]{\includegraphics[width=.5\columnwidth]{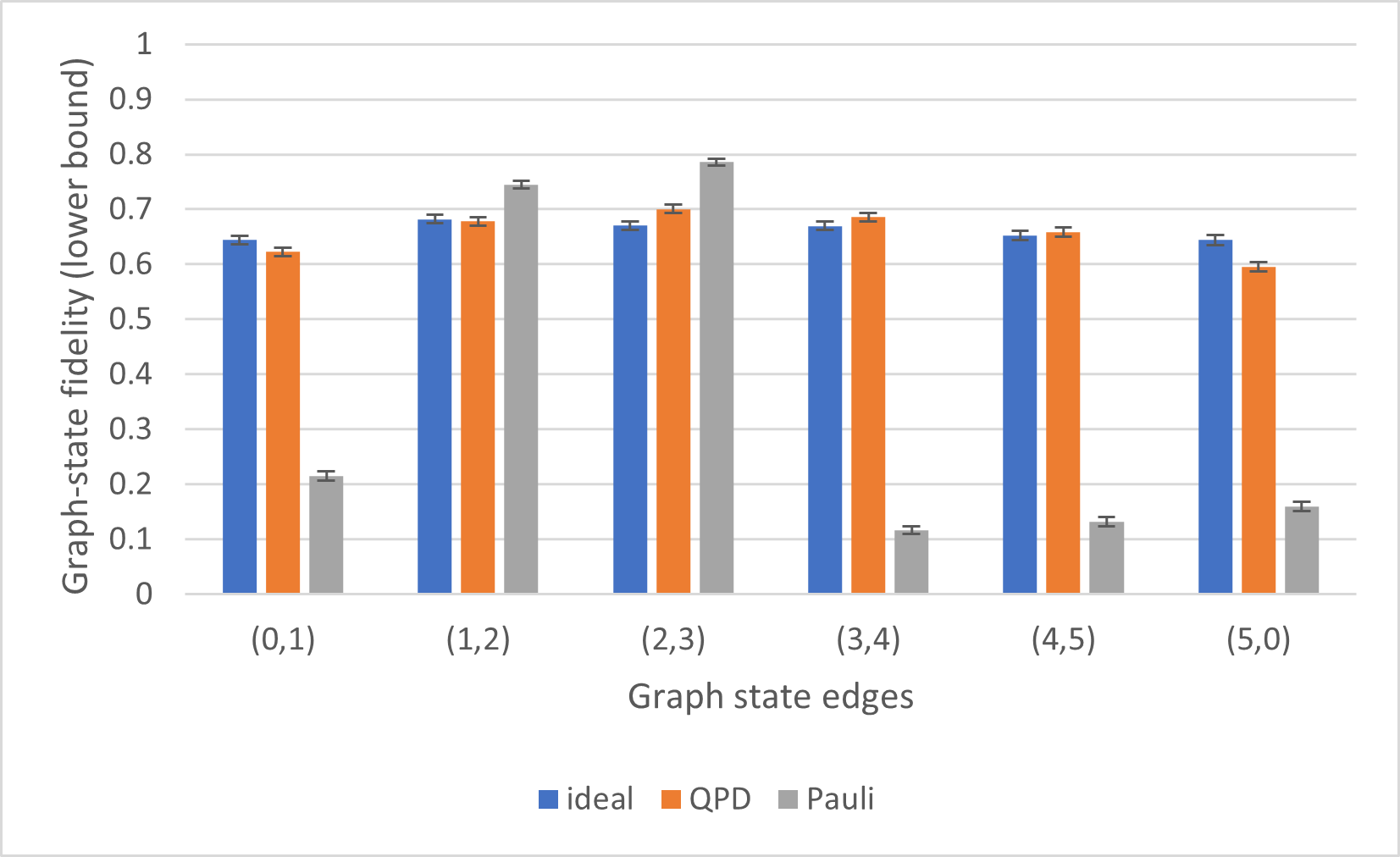} \label{fig:real_standard_fidelity}}\\
\caption[99.6\% ZALM fidelity on hardware] {\textbf{99.6\% ZALM fidelity on hardware} Individual graph state stabilizer values and a lower-bound for pair-wise graph state fidelities. For both stabilizers and fidelity, a unity value indicates perfect entanglement with no errors. }
\label{fig:real_standard}
\end{figure}

Fig.~\ref{fig:real_standard} presents the hardware results for the output stabilizers and lower-bounded fidelities when the depolarized runs are configured to emulate a 99.6\% ZALM-generated fidelity. The error bars indicate shot-based statistical uncertainty. We evaluate both the QPD and Pauli depolarizing channels described in~\cite{tittelbaugh2025modeling}, along with the ideal case in which no depolarization is applied. As discussed in Section~\ref{sec:hardware}, the unitary depolarization method requires greater qubit connectivity than is available on current hardware, and is, therefore, not included here.

Fig.~\ref{fig:real_standard_stabilizers} shows that the Pauli stabilizers are inconsistent with both expected behavior and simulation. This discrepancy arises because transpiling the Pauli depolarizing channel to IQM Emerald's native gates requires additional Hadamard operations for the $Y$ and $Z$ implementations, resulting in unequal gate counts and accumulated error across Pauli operators. The implemented operation therefore no longer represents a true depolarizing channel, as confirmed by Fig.~\ref{fig:real_standard_fidelity}, where many Pauli-depolarized fidelities fall below 50\%.

In contrast, the QPD method performs considerably better, though still below ideal for qubits on the cut (0 and 5), suggesting depolarization impacts teleportation-based protocols more strongly on real hardware than simulations predict. Overall, edge fidelity trends are consistent with Carrera Vazquez \textit{et~al.}~\cite{CarreraVazquez2024} and simulation: the cut edge shows the lowest fidelity, with adjacent edges moderately degraded. Hardware stabilizers and fidelities are uniformly lower than simulated values due to additional transpilation gates, noise processes not captured by the noise model, and the architectural mismatch between the simulated IBM Torino processor and the
  IQM Emerald hardware.

\subsection{Network Distributed Fidelity Sweep}
\label{sec:fidelity sweep}
\subsubsection{QPD Depolarization}
Consistent with Figs.~\ref{fig:sim_standard} and \ref{fig:real_standard}, qubits along cut edges and adjacent edges show decreased fidelities, with the gap diminishing as network-distributed fidelity increases. This suggests that as the fidelity of the pre-shared entanglement increases, the characteristics of the cut edge approach those of natively generated edges within a quantum computer.

\begin{figure}[htpb]
\centering
% Row 1
\subfloat[Simulated graph state vs.~distributed fidelities under QPD depolarization and a 10~ns classical communication delay.]
{\includegraphics[width=0.48\textwidth]{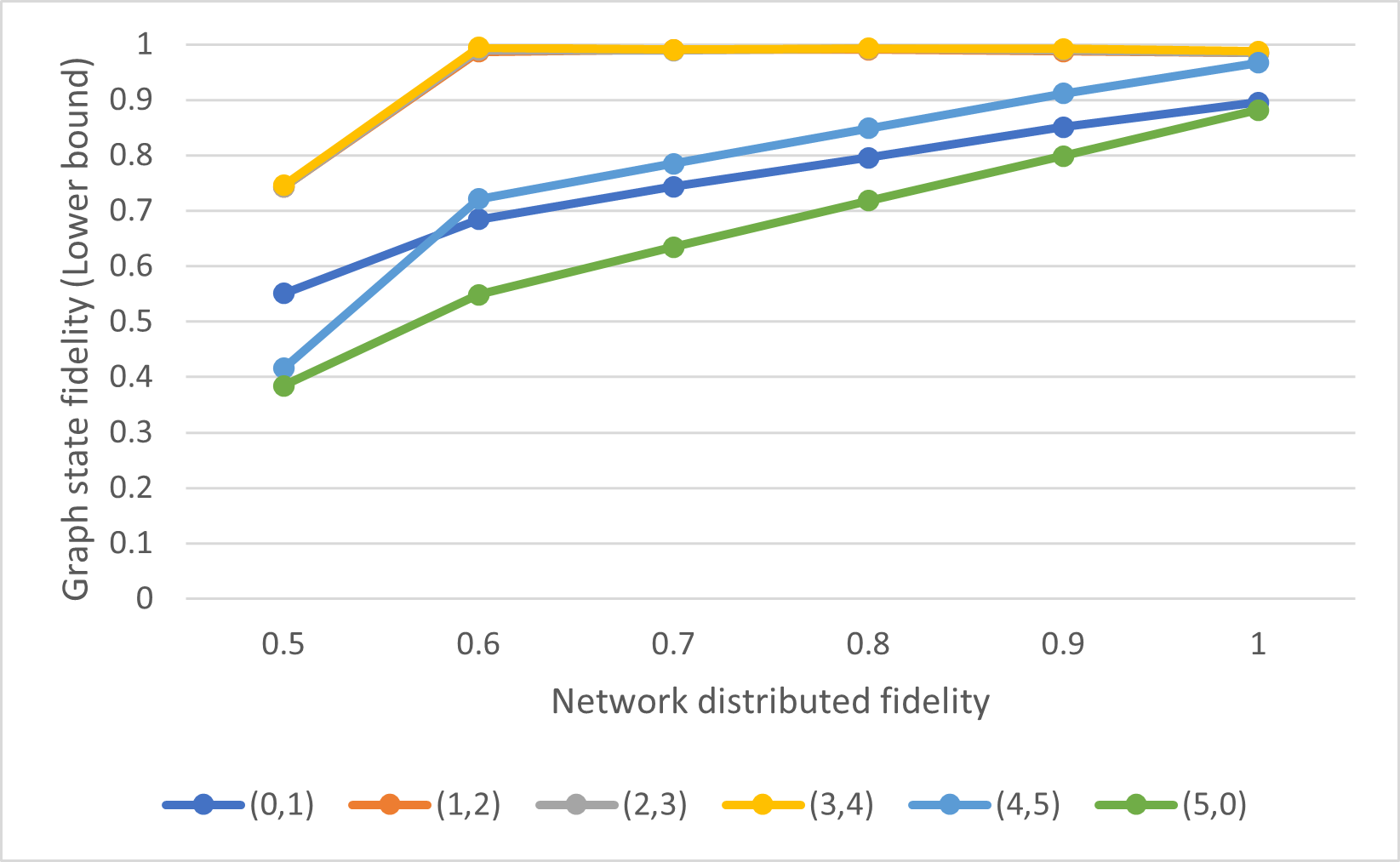}
\label{fig:QPD_sim_fidelity_sweep_fidelity}} 
\hfill
\subfloat[Graph state vs.~distributed fidelities under QPD depolarization on hardware.]
{\includegraphics[width=0.48\textwidth]{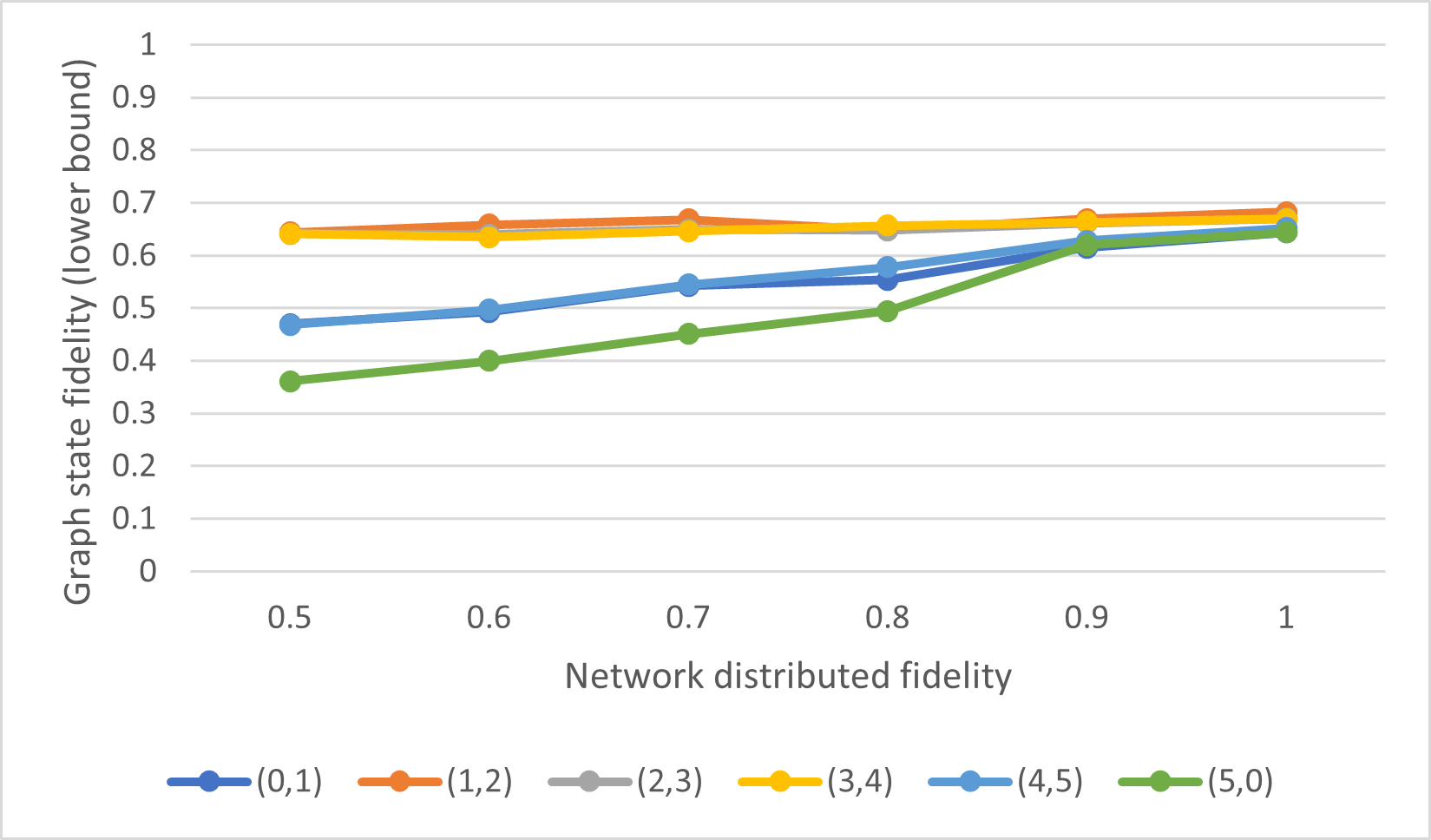}%
\label{fig:QPD_real_fidelity_sweep_fidelity}} \\
% Row 2
\subfloat[Simulated graph state vs.~distributed fidelities under Pauli depolarization and a 10~ns classical communication delay.]
{\includegraphics[width=0.48\textwidth]{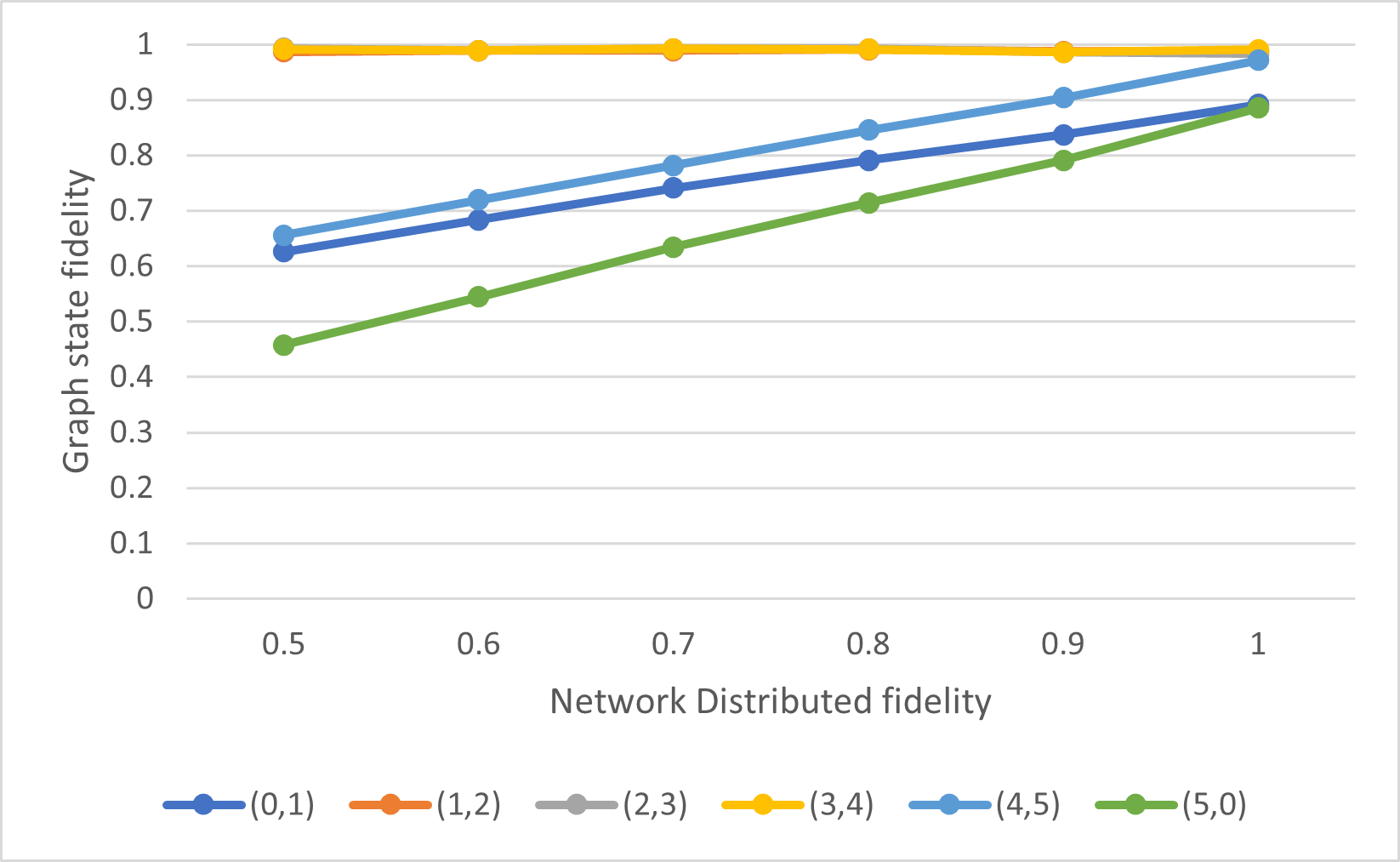}%
\label{fig:Pauli_sim_fidelity_sweep_fidelity}} 
\hfill
\subfloat[Simulated graph state vs.~distributed fidelities under unitary depolarization and a 10~ns classical communication delay.]
{\includegraphics[width=0.48\textwidth]{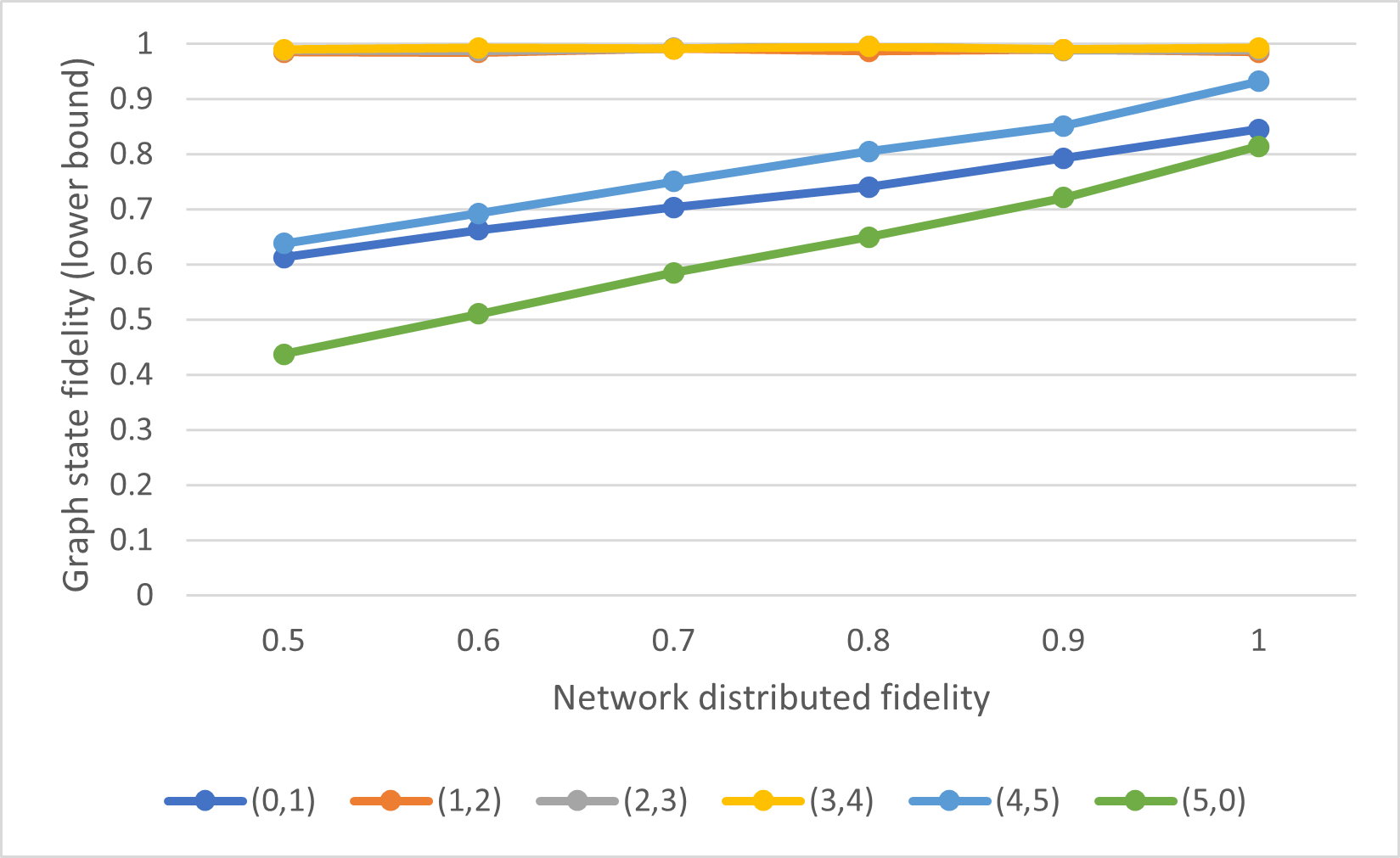}%
\label{fig:Unitary_fidelity_sweep_fidelity}} \\[1ex]
%\\[1ex]
\caption[Fidelity sweep across depolarization methods]
{\textbf{Fidelity sweep across depolarization methods} 
Subfigures (a)--(b) show the lower-bounded graph state fidelities under QPD depolarization for simulation (10~ns delay) and hardware, respectively. Subfigure (c) shows the Pauli depolarization simulation, and subfigure (d) shows the unitary depolarization simulation. Error bars are negligible.}
\label{fig:QPD fidelity sweep}
\end{figure}

In the simulated results, shown in Fig.~\ref{fig:QPD_sim_fidelity_sweep_fidelity}, the network distributed fidelity of 0.5 corresponds to a maximally mixed Bell pair, which carries no entanglement; the poor performance at this fidelity is therefore a physical expectation, not a noise artifact.

Stabilizer values and fidelities increase approximately linearly with network-distributed fidelity in both simulation and hardware, though on hardware the rate of increase lessens above 0.9, likely as internal noise becomes dominant. As discussed in~\cite{jungnitsch2011entanglement,tittelbaugh2025modeling}, entanglement is only detected at graph state fidelities greater than 0.5, which corresponds to a negative witness value. We define the creation of a graph state as successful if all edge pairs are above this value. Fig.~\ref{fig:QPD_sim_fidelity_sweep_fidelity} shows that, for our simulated results, this occurs at a network distributed fidelity of 0.6, while on the quantum computer shown in Fig.~\ref{fig:QPD_real_fidelity_sweep_fidelity}, this is only seen at network distributed fidelities of 0.9 and 1. This shows that the more error prone the QPUs, the higher quality the distributed entanglement will have to be. 

\subsubsection{Pauli Depolarization}
The Pauli depolarization method exhibits similar trends in simulation (Fig.~\ref{fig:Pauli_sim_fidelity_sweep_fidelity}), with graph state fidelities increasing linearly with network-distributed fidelity, and successful graph-state creation at fidelities of 0.6 and above. On hardware, however, the Pauli method fails to implement a true depolarizing channel, as discussed in Section~\ref{sec:hardware results}, and changing the input fidelity has no notable effect on stabilizers or pairwise fidelities.

\subsubsection{Unitary Depolarization}
The unitary depolarization method (Fig.~\ref{fig:Unitary_fidelity_sweep_fidelity}), available only in simulation due to hardware connectivity constraints, shows the same linear trend but with lower absolute fidelities, consistent with the additional gate errors shown in Fig.~\ref{fig:sim_standard}. Graph-state creation succeeds at network-distributed fidelities of 0.6 and above.

\subsection{Classical Communication Delay Sweep}
\label{sec:classical comms}
\begin{figure}[htpb]
\centering
% Row 1
\subfloat[Graph-state fidelities under no (ideal) depolarization]
{\includegraphics[width=0.48\textwidth]{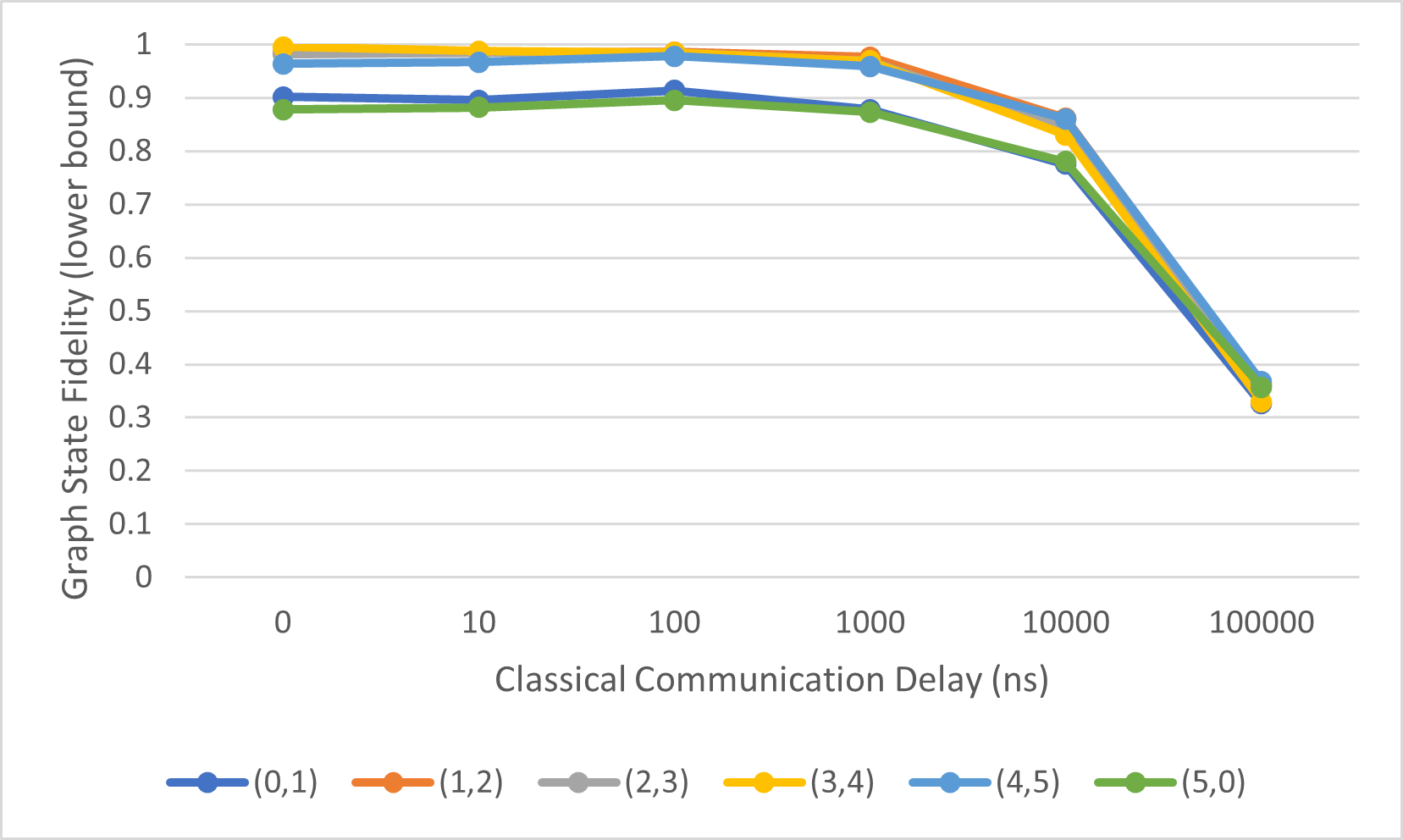}%
\label{fig:Delay_ideal_fidelity}} 
\hfill
\subfloat[Graph-state fidelities under QPD depolarization]
{\includegraphics[width=0.48\textwidth]{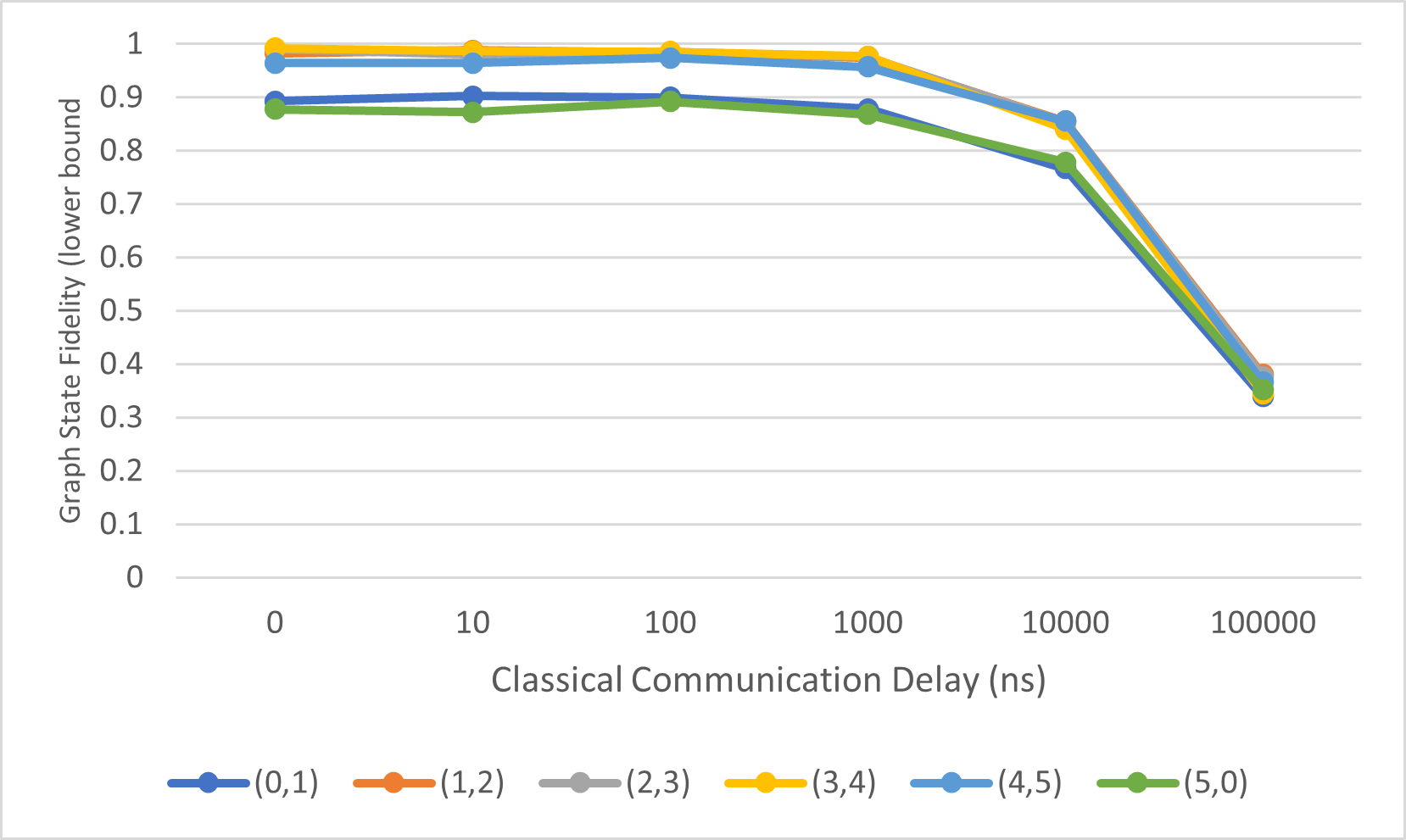}%
\label{fig:Delay_QPD_fidelity}} \\
\vspace{-0.8em}
\caption[Classical delay sweep]
{\textbf{Classical delay sweep} 
Logarithmic simulated classical delay sweep assuming a 99.6\% network distributed fidelity.}
\label{fig:Classical Communication Delay sweep}
\end{figure}

Fig.~\ref{fig:Classical Communication Delay sweep} shows the ideal and QPD depolarization cases under varying classical communication delay, simulated at 99.6\% ZALM fidelity with thermal relaxation noise proportional to delay over ideal optical fiber. Consistent with previous results, the cut edge~(0,5) yields the lowest fidelity, with adjacent edges moderately degraded. The Pauli and unitary methods (not shown) exhibit similar trends with slightly lower absolute fidelities.

Our results show that for classical communication delays between 0 and 100~ns, there is no appreciable loss in stabilizer values or graph state fidelity. This is encouraging for data-center-scale distributed quantum computing, where interconnects are typically on the order of a few meters or less of optical fiber. A small but noticeable decrease in both stabilizer values and fidelities appears around 1,000~ns, corresponding to approximately two hundred meters of transmission distance. Even so, all depolarization methods maintain fidelities above the 50\% entanglement threshold up to delays of approximately 10,000~ns, or a few kilometers. Only at 100,000~ns, corresponding to tens of kilometers of fiber, does thermal relaxation noise become dominant, reducing the resulting graph state fidelities below the entanglement threshold. 

\section{Conclusion}
\label{sec:conc}
Overall, this work successfully created and demonstrated a framework for modeling the effects of network-distributed entanglement on creating a virtual entanglement graph state on a quantum computer, an essential step to enabling distributed quantum computing. This is based on the results in  Carrera Vazquez \textit{et al.}, where they show that this methodology can be extended across multiple QPUs~\cite{CarreraVazquez2024}. We model the decreased fidelity entangled pairs produced by the ZALM entanglement process with three distinct descriptions of depolarizing channels, an extension of the QPD used in  Carrera Vazquez \textit{et al.}, randomly applying Pauli error, and a Stinespring dilated unitary operator. These methods are examined in simulation, with the QPD and Pauli methods additionally evaluated on quantum hardware. Throughout the tests we found that the QPD method consistently performed the best of the proposed methods, at the cost of more runs in both simulation and on the quantum computer. This advantage is expected, as QPD introduces no additional gates; the more significant finding is that mathematically equivalent noise implementations diverge substantially on real hardware.

We discovered that to effectively create entanglement, i.e., have a graph state fidelity of greater than 50\%, on the quantum hardware the network-distributed entanglement fidelity needs to be quite high, 90\% or greater. Finally, we examined the effect of classical communication delay on the creation of graph states. We showed that the increase of thermal relaxation noise due to the extended delay is negligible over tens of meters, a promising result for data centers that often only need a few meters of fiber to connect multiple QPUs. In fact, we find that only at tens of kilometers do the delays prevent the teleportation of entanglement in the resulting graph state. 

%\subsection{Future Work} -> paragraph?
The results here are influenced by transient hardware noise; future work could improve statistical robustness by averaging over multiple calibration cycles. We also plan to benchmark our virtual Bell pair against a physical Bell pair prepared with a true entangling gate on the same hardware. Error-mitigation techniques such as TREX and optimized qubit mapping~\cite{CarreraVazquez2024} could be incorporated to improve hardware results. Extending this framework to two cut Bell pairs connecting distinct graph states, and to algorithms using graph states as computational resources such as the Steane code~\cite{steane1996multiple}, would provide a more complete picture of network-enabled distributed quantum computing.

\begin{credits}
\subsubsection{\ackname} This work is supported by the National Science Foundation under Grant No. CNS-2107265 and Research Ireland under grants 20/US/3708, 21/US-C2C/3750, and 13/RC/2077 P2.

\subsubsection{\discintname}
The authors have no competing interests to declare that are relevant to the content of this article. 
\end{credits}
\bibliographystyle{splncs04}
\bibliography{references}

%\end{thebibliography}
\end{document}